# Radiative transfer approach using Monte Carlo Method for actinometry in complex geometry and its application to Reinecke salt photodissociation within innovative pilot-scale photo(bio)reactors.

V. Rochatte[1,2], G. Dahi[1,2] A. Eskandari[1,2], J. Dauchet[3,2], F. Gros[3,2], M. Roudet[3,2], J.-F. Cornet[3,2]

(1) Université Clermont Auvergne, Université Blaise Pascal, Institut Pascal, BP 10448, F-63000 CLERMONT-FERRAND, FRANCE
(2) CNRS, UMR 6602, IP, F-63178 Aubière, FRANCE
(3) Université Clermont Auvergne, SIGMA Clermont, Institut Pascal, BP 10448, F-63000 CLERMONT-FERRAND, FRANCE

**Astract**

In this article, a complete radiative transfer approach for estimating incident photon flux density by actinometry is presented that opens the door to investigation of large-scale intensified photoreactors. The approach is based on an original concept: the analysis of the probability that a photon entering the reaction volume is absorbed by the actinometer. Whereas this probability is assumed to be equal to one in classical actinometry, this assumption can no longer be satisfied in many practical situations in which optical thicknesses are low. Here we remove this restriction by using most recent advances in the field of radiative transfer Monte Carlo, in order to rigorously evaluate the instantaneous absorption-probability as a function of conversion. Implementation is performed in EDStar, an open-source development environment that enables straightforward simulation of reactors with any geometry (directly provided by their CAD-file), with the very same Monte Carlo algorithm. Experimental investigations are focused on Reinecke salt photodissociation in two reactors designed for the study of natural and artificial photosynthesis. The first reactor investigated serves as reference configuration: its simple torus geometry allows to compare flux densities measured with quantum sensors and actinometry. Validations and analysis are carried out on this reactor. Then, the approach is implemented on a 25 L photobioreactor with complex geometry corresponding to one thousand light-diffusing optical fibers distributing incident photons within the reaction volume. Results show that classical actinometry neglecting radiative transfer can lead to 50 percent error when measuring incident flux density for such reactors. Finally, we show how this radiative transfer approach paves the way for analyzing high conversion as a mean to investigate angular distribution of incident photons.

**Highlights:**

A novel and improved extent of actinometry to determine photon flux is presented.

Latest advances in Monte Carlo Method for radiative transfer have been used.

Photon absorption probability by the actinometer is defined as a new tool.

Complex geometries, pilot plant photo(bio)reactors can now be easily addressed.

**Keywords:**

Actinometry, radiative transfer, Monte Carlo method, complex geometry, photon absorption probability

## 1. Introduction

Photons are known to generate biochemical or chemical changes in living or non-living systems. Such an ability is called actinism [1] as defined in the 19[th] century [2]. The metering of actinism was under scope in the early 20[th] at the beginning of photochemistry [3] and among all the parameters that are investigated (light source, vessels, etc.), the estimation of photon flux entering the reactive medium is still of prime importance since this information is the start of basis of any subsequent study of the photochemical system. To achieve this assessment various means can be used such as bolometers, photodetectors…[4] and besides those physical apparatus, chemical actinometers developed in the 20[th] century, like uranyl oxalate [5] or potassium ferrioxalate [6], are nowadays widely used for the determination of photon flux entering small scale vessels [7]. These actinometers undergo a light-induced wavelength dependent reaction for which a quantum yield, is accurately known [8] in a given spectral range. This quantum yield being established, measuring the reaction rate allows the estimation of the volumetric rate of absorbed photon, $\mathcal{A}$, or even more interesting the determination of the hemispherical photon flux density entering the photoreactor, $q_\cap$. Practically, the classical operating conditions (high optical thickness, low conversion ratio, *i.e.* short reaction time) [9] are meant to ensure that all the emitted photons are absorbed by the actinometer, leading to a remarkably simple treatment of the experiment: the temporal evolution of the concentration is then linear with a slope proportional to $q_\cap$, that is easily modelized without the need of advanced considerations on the radiative transfer within the reaction volume (in that case, the only radiative consideration is that all the photons are absorbed). The simplicity of the implementation and the accuracy of the obtained results made actinometry very popular, such that the scope of its application was called upon to extend.

Indeed, frequently encountered are situations where it is not possible to ensure that all the emitted photons are absorbed by the actinometer; models including radiative transfer (for the estimation the local volumetric rate of energy absorption, $\mathcal{A}$, (LVREA)) have been introduced for this purpose [10], with a recent extension by Zalazar et al. to systems involving light scattering, absorption by reaction products and polychromatic light source [11]. However to the best of our knowledge the implementation of actinometry in situations with partial absorption is to the date limited to one-dimensional geometries at small scales, that is not compatible with our studies of intensified processes, at pilot plant scale, with high-volume and surface productivities enabled by the dilution of the incident radiation within the reaction volume thanks to an optimal spatial arrangement of light-diffusing optical fibers [12].

The purpose of the present article is therefore to present a general methodology for estimating incident photon flux density by actinometry, considering incomplete light absorption, based on the photon absorption probability by the actinometer, p, and not on the radiation field, from a thorough characterization of an actinometer to a rigorous radiative transfer description and resolution in complex geometries. The first part of this article will present briefly the classical modelling of actinometry (*i.e.* linear temporal evolution) that can be usually found in literature. In the second section our general methodology, using radiative transfer approach, for the estimation of the absorption probability, p, will be presented. In addition to the presentation and full characterization of the actinometric system utilized, our radiative transfer approach for the estimation of p will be employed and compared (for validation purpose) with the analytical expressions in simple monodimensionnal geometry at lab scale. The extension of our methodology to a 3D complex geometry at a pilot scale will be made in the last part of this paper.

## 2. Model for actinometry

### 2.1 Kinetic model

Typical phototransformation reaction used in actinometry can be summarized as follows:

$$A \xrightarrow{h\nu} B + C \qquad (1)$$

where photon absorption by the actinometer A (here noted hν) leads to the production of two chemical species, or pseudo-species, B and C. Hereafter focus is on the representative situation where C absorbs photons in the same wavelength range as A; on the contrary of B, that doesn't absorb photons. Actinometry experiments are usually conducted within batch reactors and the extent of reaction is monitored by titration of either A, B or C (most of the time B).

The modelling of the experiment starts with the formulation of the mass balance equation on the reactant A within the batch reactor:

$$\frac{dC_A}{dt} = \langle r_A \rangle \qquad (2)$$

where $C_A$ is the concentration of reactant A (mol.m$^{-3}$). Then, phototransformation kinetics gives the expression of the mean volumetric reaction rate $\langle r_A \rangle$ in the reaction volume as a function of photon absorption rate:

$$\langle r_A \rangle = -\Phi \langle \mathcal{A} \rangle \tag{3}$$

with $\langle \mathcal{A} \rangle$, the mean volumetric rate of photon absorption (MVREA) by the actinometer A (a mean spectrally and volumetrically averaged quantity of $\mathcal{A}$, expressed in $\mu mol_{hv}.m^{-3}.s^{-1}$) and $\Phi$, the quantum yield of the photoreaction. According to the definition of actinometers, the kinetic coupling with photon absorption is linear, that is to say that $\Phi$ is independent of $\mathcal{A}$, so the mean spatial integration of the volumetric local rate gives Eq. (3). Moreover, in the following that $\Phi$ is assumed to be independent of the radiation wavelength, as it is the case for many (but not all) actinometers [4].

**2.2 New formulation of actinometry using the proportion of absorbed photons, p**

Measuring incident photon flux density $q_\cap$ (expressed in $\mu mol.m^{-2}.s^{-1}$) with actinometry requires constructing a radiative transfer model for the rate of photon absorption $\langle \mathcal{A} \rangle$ as a function of $q_\cap$. Here we argue that conceptual and practical benefits can be expected from a formulation of $\langle \mathcal{A} \rangle$ in terms of the probability that a photon entering the system will be absorbed by the actinometer A. In order to establish such a formulation, let us introduce the area $S_{light}$ of the photon emitting surface, which is here assumed to ensure a mean homogeneous incident surface density flux $q_\cap$. Moreover we assume that photon emission due to fluorescence within the volume can be neglected[1]. Therefore, multiplying the photon flux $S_{light}q_\cap$ entering the reactive medium by the probability p that such a photon is absorbed

---

[1] Luminescent emission has a significant influence on $\langle \mathcal{A} \rangle$ only if, first, spectral range of luminescent radiation can lead to phototransformation of the actinometer A and second, if absorption rate of luminescent photons by A cannot be neglected compared with direct absorption of photons emitted at the reactor surface. Overall, luminescence can generally be neglected when evaluating $\langle \mathcal{A} \rangle$.

by A leads to the overall rate of photon absorption and dividing it by the volume V of the medium gives:

$$\langle \mathcal{A} \rangle = \frac{S_{light} \cdot q_\cap}{V} p \qquad (4)$$

where $\frac{S_{light} \cdot q_\cap}{V}$ is the maximum value for $\langle \mathcal{A} \rangle$. In Eq. 4, $p \in [0,1]$ is the probability that a photon entering the reactive medium is absorbed by the actinometer A. Note that, as an important property, p is independent of the incident flux density $q_\cap$.

Substituting Eq. 4 into Eq. 3 and replacing the resulting expression back into Eq. 2, the mass balance equation on reactant A becomes:

$$\frac{dC_A}{dt} = -\Phi \cdot a_{light} \cdot q_\cap \cdot p(t) \qquad (5)$$

where we introduced the specific illuminated surface $a_{light} = \frac{S_{light}}{V}$ that is a key engineering parameter for the optimisation of photo-reactive systems.

This new formulation in terms of the proportion p of absorbed photons permits to distinguish the influence of incident flux density $q_\cap$ (photon absorption rate is proportional to $q_\cap$ due to linearity of the radiative transfer equation) and the influence of photon transport that is contained in the value of p only. Within the frame of this formulation, the question of the radiative transfer modelling comes down to the evaluation of p, that is a complicated function of chemical species concentrations and reactor geometry, but that is independent of the incident flux density $q_\cap$. In a practical point of view, this formulation allows for pre-tabulation of p values.

### 2.3 Limitations of classical actinometry

As stated earlier if all emitted photons are steadily absorbed by the actinometer A, *i.e.* p = 1, the right hand side of Eq. 5 is constant hence leading to the widespread linear temporal evolution of the concentration [8] with a slope proportional to $q_\cap$:

$$C_A = C_{A_0} - \Phi \cdot a_s \cdot q_\cap \cdot t \qquad (6)$$

However, p could be different from 1 and time-variable when:

1) the absorption optical thickness, $e = C_A E_A L$, of the solution is not high enough hence a high number of photons goes through the solution without being absorbed. It can be found when the actinometer concentration (at the beginning of the reaction or for high conversion) and/or vessel thickness and/or mean absorption coefficient are low for a given wavelength.

2) the actinometer is not the only absorbing specie in solution hence photons are likely to be lost due to the absorption by the other chemical species present in solution (for example photochemical reaction products C in Eq. 1).

Either 1) and/or 2) situations can be found in technological objects we develop for research (photoreactor or photobioreactor [12]) as innovative future solar photoprocesses with high thermodynamic efficiency. But the both cases are encountered for high conversion where actinometer concentration, $C_A$, is small and absorbing product concentration, $C_C$, is high. In these cases Eq. 6 can no longer be applied. To correctly use the results of actinometry experiment in all these situations, it is therefore compulsory to solve the radiative transfer equation within the photoreactor [11] to estimate p.

## 2.4 Radiative transfer description for actinometry for the estimation of p

In this section, focus is made on the calculation of the proportion p of photons absorbed by the actinometer A, which is needed when analysing actinometry experiments. A Monte Carlo algorithm is presented that evaluates the value of p obtained by rigorously solving the radiative transfer equation for any photoreactor design and any reaction extent. Numerical implementation of this algorithm is discussed in Section 2.5 and summarised in Fig. 1.

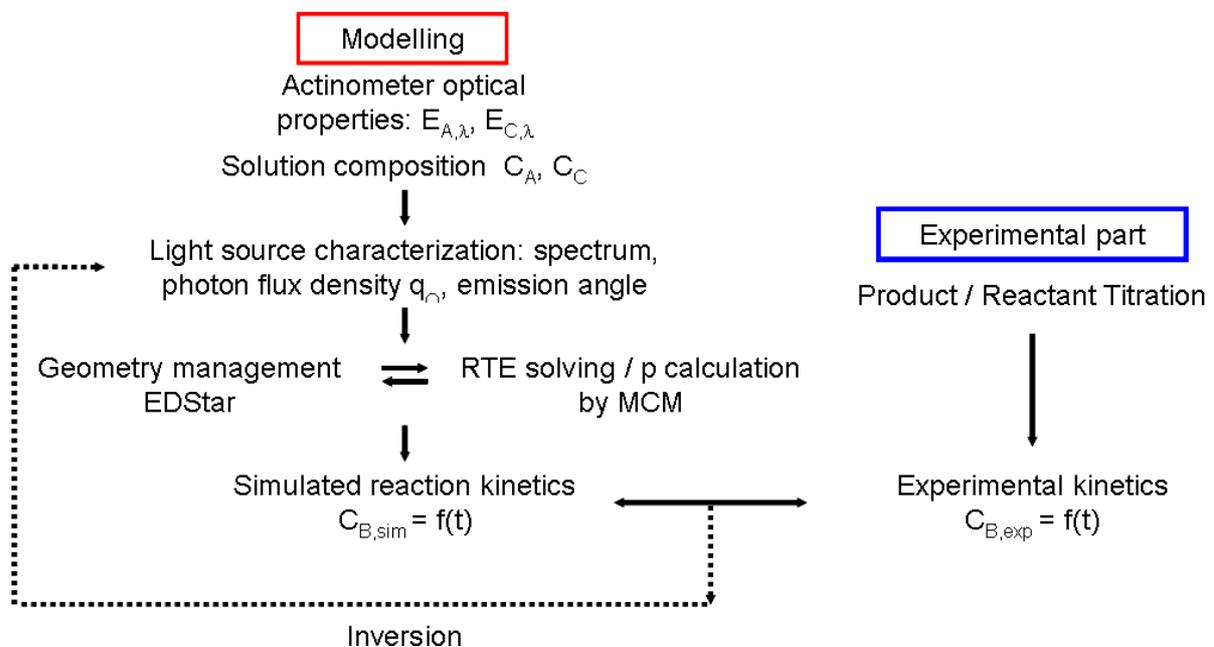

Fig. 1. Summary of the procedure for photon flux determination.

Obtaining the proportion of absorbed photons p as a function of reaction extent requires solving the steady-state radiative transfer equation for spectrally absorbing, non-emitting and non-scattering medium (steady-state is considered for radiative transfer because the radiative time constant is much shorter [13] than the chemical reaction time constant [14]).

The reactor is assumed to be perfectly stirred, with homogeneous concentrations for the absorbing chemical species A and C. Therefore, the absorption coefficient within the reaction volume V is:

$$k_\lambda = C_A E_{A,\lambda} + C_C E_{C,\lambda} \tag{7}$$

where $C_A$ and $C_C$ are respectively the actinometer A and absorbing by-product C concentrations (in mol.m$^{-3}$); $E_{A,\lambda}$ and $E_{C,\lambda}$ are respectively the molar absorption cross-section of A and C (in m$^2$.mol$^{-1}$). When a photon is absorbed within the medium, $\frac{C_A E_{A,\lambda}}{k_\lambda}$ is the probability that the photon is absorbed by the actinometer; $\frac{C_C E_{C,\lambda}}{k_\lambda} = 1 - \frac{C_A E_{A,\lambda}}{k_\lambda}$ is the probability that the photon is absorbed by the by-product. Usually by-product absorption is not considered nor treated in the literature ($E_{C,\lambda}$ is set equal to 0), but experimental results presented in section 3 will indicate that neglecting this phenomenon can lead to significant errors.

The boundary conditions associated with the radiative transfer equation are the following:

- The reaction volume V is bounded by a given surface S (the geometry of S depends on the studied reactor).

- Reflectivity ρ(**x**) and distribution $p_\Omega^\rho(\boldsymbol{\omega}, \mathbf{x})$ of the reflection directions **ω** are known at any location **x** of S (specular, diffuse, etc. depending on the materials).

- The part of S that is emitting photons is noted $S_{light}$. The emission spectrum $p_\Lambda^{light}(\lambda)$ of the source and the distribution $p_{\Omega_0}^{light}(\boldsymbol{\omega_0}, \mathbf{x_0})$ of incident directions $\boldsymbol{\omega_0}$ are known at any location $\mathbf{x}_0$ of $S_{light}$ (collimated, diffuse, etc. depending on the light source).

These boundary conditions depend on the studied photoreactor; two practical examples will be discussed in section 3 (a quasi-one-dimensional torus reactor and a real-world pilot-plant photobioreactor with complex geometry for S).

Now that the problem is well-posed, let us focus on the evaluation of p. Monte Carlo Method (MCM) is the standard method to obtain reference solution of the radiative transfer equation [13] [15]. For examples, Moreira et al. [16] and [17] studied radiative transfer in a slurry of $TiO_2$ particles with this method in a cylindrical photoreactor. In addition, on the basis of its integral formulation [18], Cassano's team has also used MCM to solve the radiative transfer equation in complex geometry within the context of photoreaction [19].

Here we design a Monte Carlo algorithm that evaluates the proportion of photon absorbed by the actinometer A, using most recent advances in the field of radiative transfer Monte Carlo along the line of [20] and [21]. The algorithm consists in statistically sampling optical paths (that is to say "photon trajectories") by simulating statistical physics of transport for photon emission, reflection and absorption, in a quite intuitive manner. For each optical path j, a weight $w_j$ is retained: $w_j$=1 if the photon is absorbed by the actinometer A; $w_j$=0 if the photon is absorbed by the by-product C, or lost at a surface of the reactor (losses include dissipation and transmission). Repeating N times the sampling procedure (j=1,2,...,N), an estimator $\hat{p}_N$ of p is constructed as the average of the Monte Carlo weights:

$$p \approx \hat{p}_N = \frac{1}{N}\sum_{j=1}^{N} w_j \qquad (8)$$

and the numerical error for this estimation is given by the standard deviation:

$$\sigma_N = \sqrt{\frac{\frac{1}{N}\sum_{j=1}^{N} w_j^2 - \hat{p}_N^2}{N-1}} \qquad (9)$$

The optical-path sampling procedure is the following[2]:

1. An emission location $\mathbf{x_0}$ is uniformly sampled[3] over the surface $S_{light}$ and an emission direction $\boldsymbol{\omega_0}$ is sampled over the inner hemisphere at $\mathbf{x_0}$, according to the distribution $p_{\Omega_0}^{light}(\boldsymbol{\omega_0})$.

2. A wavelength $\lambda$ is sampled over the spectral range $[\lambda_{min}, \lambda_{max}]$ of the source, according to the emission spectrum $p_\Lambda^{light}(\lambda)$. This wavelength sets the value of the molar absorption cross-sections $E_{A,\lambda}$ and $E_{C,\lambda}$ and the absorption coefficient $k_\lambda$ for the current optical path.

3. An absorption (as a reminder, the medium is non scattering) length $l_0$ is sampled over $[0,+\infty]$ according to the exponential extinction law $k_\lambda \exp(-k_\lambda \cdot l_0)$. Now that $\{\mathbf{x_0}, \boldsymbol{\omega_0}, l_0\}$ have been sampled, the first interaction location $\mathbf{x_1}$ is determined. As discussed in the following section, pure geometrical considerations are easily translated into scientific computation libraries. For a given couple $\{\mathbf{x_0}, \boldsymbol{\omega_0}\}$ such libraries provide us with the location $\mathbf{y}$ of the first time the half-line starting at $\mathbf{x_0}$ in the direction $\boldsymbol{\omega_0}$ intersects the bounding surface S. If the distance to the bounding surface $||\mathbf{y} - \mathbf{x_0}||$ is smaller than the absorption length, the optical path interacts with the surface, otherwise absorption occurs inside the reaction volume :

$$\mathbf{x_1} = \begin{cases} \mathbf{y} & \text{if } \|\mathbf{y} - \mathbf{x_0}\| < l_0 \\ \mathbf{x_0} + l_0 \cdot \boldsymbol{\omega_0} & \text{otherwise} \end{cases}$$

---

[2] For a better understanding, this algorithm is schematically represented in Fig. 8 in the case of a complex geometry.
[3] Photon locations $\mathbf{x}$ are expressed as coordinates $\mathbf{x} = (x_1, x_2, x_3)$.

Then, a branching test is performed depending on the interaction location:

- In case of an interaction with the surface S (i.e. $\mathbf{x_1} \in S$), a Bernoulli trial is performed: a random number $r_1$ is uniformly sampled over the unit interval; if $r_1$ is lower than the reflectivity $\rho(\mathbf{x_1})$ – the optical path is reflected - then a reflection direction $\boldsymbol{\omega_1}$ is sampled according to the material's bidirectional reflection distribution function $p_\Omega^\rho(\boldsymbol{\omega_1}, \mathbf{x_1})$ and the algorithm loops to step 3 (the indexes being incremented); if $r_1$ is greater than the reflectivity $\rho(\mathbf{x_1})$ – loss at the surface occurs – then the optical path sampling procedure is terminated and the Monte Carlo weight for this path is set equal to 0.

- In case of an interaction within the volume V (i.e. $\mathbf{x_1} \in V$), a Bernoulli trial is performed: a random number $r_1$ is uniformly sampled over the unit interval; if $r_1$ is lower than the probability $\frac{C_A E_{A,\lambda}}{k_\lambda}$ absorption by the actinometer A occurs, thus the optical path sampling procedure is terminated and the Monte Carlo weight for this path is set equal to 1; if $r_1$ is greater than the probability $\frac{C_A E_{A,\lambda}}{k_\lambda}$ absorption by the by-product C occurs, thus the optical path sampling procedure is terminated and the Monte Carlo weight for this path is set equal to 0.

In addition to its intuitive features, this Monte Carlo algorithm rigorously evaluates p. Indeed, the above sampling procedure is the strict algorithmic translation of the integral formulation for p, obtained by formally solving the radiative transfer equation [20], [21]:

$$p = \int_S dx_0 \frac{1}{S} \int_{2\pi^+(x_0)} d\varpi_0 p^{light}_{\Omega_0}(\varpi_0, x_0) \int_{\lambda_{min}}^{\lambda_{max}} d\lambda p^{light}_\Lambda(\lambda) \int_0^{+\infty} dl_0 k_\lambda \exp(-k_\lambda l_0) \left[ H(x_1 \in S)\rho(x_1)I_1 + H(x_1 \in V) \frac{C_A E_{A,\lambda}}{k_\lambda} \right] \quad (10)$$

with $2\pi^+(x)$ the inner hemisphere at location x of the surface S and $I_1$ recursively defined as:

$$I_j = \int_{2\pi^+(x_j)} d\varpi_j p^\rho_{\Omega_j}(\varpi_j, x_j) \int_0^{+\infty} dl_j k_\lambda \exp(-k_\lambda l_j) \left[ H(x_{j+1} \in S)\rho(x_{j+1})I_{j+1} + H(x_{j+1} \in V) \frac{C_A E_{A,\lambda}}{k_\lambda} \right] \quad (11)$$

where the Heaviside function $H(x_j \in D)$ takes the value 1 when the condition $x_j \in D$ is satisfied and 0 otherwise.

For a simple one-dimensional configuration (see *e.g.* Fig. 3, our slab-like torus reactor) without reflection (*i.e.* ρ = 0) and with monochromatic incident radiation, Eq. 10 significantly simplifies and p becomes straightforward to evaluate analytically, as presented in appendix B. This analytical reference solution will be used in section 3.2 to validate our Monte Carlo algorithm. For the other configurations investigated in section 3, the algorithm is used to evaluate p rigorously. But first, let us present how we numerically implement this Monte Carlo algorithm within any geometry of reactor.

### 2.5 Numerical implementation

With Monte Carlo method, the difficulty associated with geometric complexity is reduced to that of calculating intersections between straight rays and the complex reactor's bounding surface S (see the algorithm presented in the previous section). This pure geometrical consideration that has no direct relation with physical reasoning can be translated into scientific computation libraries, which implies that the same optical-path sampling procedure (*i.e.* the same physics) can be implemented within any geometry of reactor, in a quite straightforward manner.

Our Monte Carlo algorithm for the estimation of p is implemented within the free EDStar development environment [21] [22] [23] that makes available scientific libraries and computation tools developed by the computer graphics research community for geometrically defining complex geometries and accelerating photon tracking in such geometries. Implementation within EDStar allows separating completely the description of the reactor's geometry from the description of the physics (that corresponds to the optical path sampling procedure accounting for emission, reflection and absorption of photons). In a practical point of view, a Computer Aided Design (CAD) file is provided that defines the surface S and its properties (emission and reflection properties corresponding to the boundary conditions of the radiative transfer equation). The Monte Carlo algorithm is programmed in a separate file where we have access to abstractions and functions that are used to code the optical path sampling procedure regardless of the geometry specified in the CAD file. For example, when programming Step 3 of the algorithm, a function available in EDStar returns the first intersection between the half line and the surface S specified in the CAD file (we have access to the location of the intersection and the reflection properties at this location). Examples of algorithms programmed within EDStar are available in [21]. This orthogonality between geometric data and sampling procedures perfectly meets the needs of our study: first the algorithm is programmed without worrying of technical aspects that have no direct relation with physical reasoning (EDStar's scientific computation libraries handle statistical treatments, parallel implementation and pure geometrical reasoning) then the algorithm is validated in a simple geometry and finally it can be directly implemented within any complex reactor geometry (without modifying the sampling procedure).

The Monte Carlo algorithm is used to tabulate the value of p as a function of the concentrations $C_A$ and $C_C$ of actinometer A and absorbing by-product C. Then, for a given flux density $q_\cap$, the differential equation Eq. 5 is numerically solved using a standard explicit Euler scheme (the value of p is interpolated within the p-table at each time step). Finally, the flux density $q_\cap$ is obtained by solving the inverse problem: we retain the value of $q_\cap$ minimizing the mean square error between experimental and predicted concentrations at different times (see Fig. 5 and 9). Minimization is performed using a simplex algorithm (fminsearch Matlab function). All these steps are summarized in Fig. 1.

## 3. Experimental application

### 3.1 Choice and characterization of the actinometer

Our researches are aimed at studying natural or artificial photosynthesis and developing associated efficient processes [12]. The wavelength range of interest is actually 350-800 nm and thus our light sources are the Sun or artificial lights in the visible spectrum. One of the most interesting wide wavelength range actinometer that can be used in these cases is the Reinecke salt actinometer as developed by Wegner and Adamson [24] and later used by other research teams ([10][25][26][27]) for photon flux estimation. It is one of the simplest chemical systems for the visible region [28] and its quantum yield is nearly constant (equal to 0.29±0.02) over a wide wavelength range. At low conversion, its decomposition corresponds to a thiocyanate ligand substitution by a water molecule (Eq. 12):

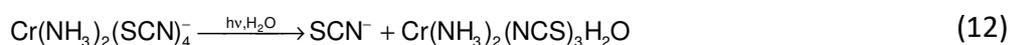

$$Cr(NH_3)_2(SCN)_4^- \xrightarrow{h\nu, H_2O} SCN^- + Cr(NH_3)_2(NCS)_3H_2O \qquad (12)$$

It could also be written as Eq. 1 (with letters used in section 2): $A \xrightarrow{h\nu} B + C$

B and C letters could respectively describe thiocyanate anion and $Cr(NH_3)_2(SCN)_3H_2O$ complex.

In this work, the Reinecke salt (supplied by Sigma Aldrich) was used according to the procedure developed by Cornet et al. [10] (with solubilisation and stripping steps different from Wegner and Adamson's initial work [24]) and using thiocyanate ion titration to follow reaction progress. The initial Reinecke salt concentrations were 15 mol.m$^{-3}$.

In addition, we determined the absorption coefficients, $E_\lambda$, of the Reinecke salt and its photolytic by-product $(Cr(NCS)_3H_2O(NH_3)_2)$ using the original work of Wegner and Adamson

[24] and our experimental results, obtained when measuring solution absorbance at low conversion in a high precision optical bench (consisting of a modified FLX-Xenius spectrofluorometer (Safas) in association with a 6 inch integrating sphere (Labsphere). The results are presented in Fig. 2 (numerical values are included in appendix A and added in the supplementary information).

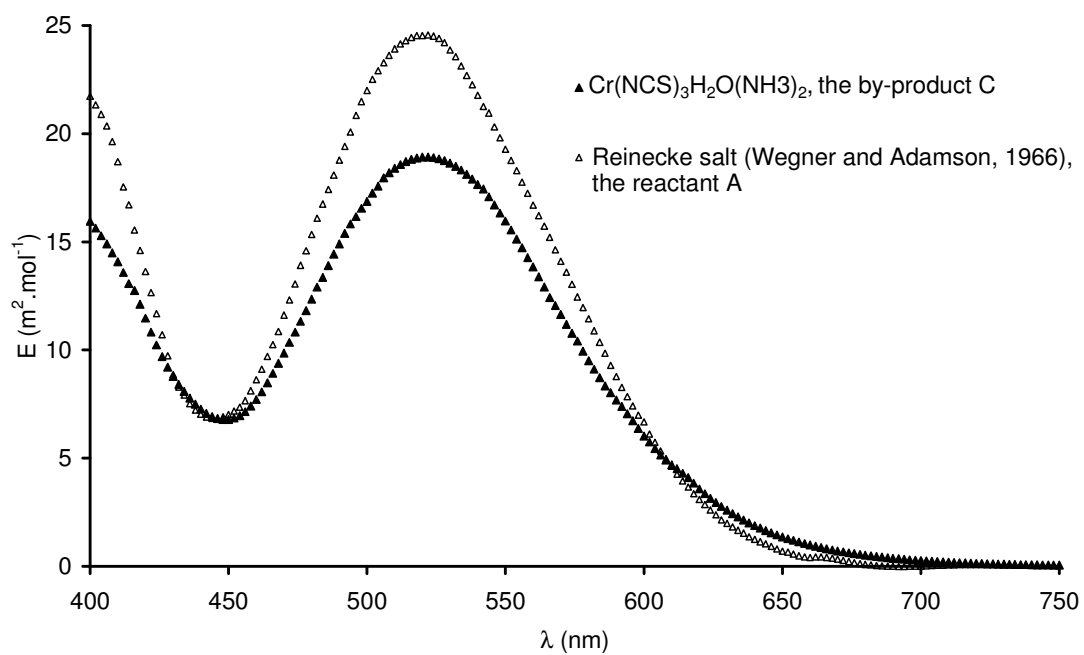

Fig. 2. Absorption coefficient of Reinecke salt and by-product against wavelength.

As can be seen, the Reinecke salt actinometer covers a wide wavelength range in the visible region on the contrary to the classical ferrioxalate actinometer (whose action spectrum is moreover limited to 450-500 nm [4], [29]). The absorption coefficients of the two chemical species are in the same order of magnitude, the photon absorption by the by-product can't be ignored (once again on the contrary to the classical ferrioxalate actinometer where the by-products are absorbing at even shorter wavelength than the reactant). The identified spectrum for the by-product is in agreement with partial literature data [30] [31] [32], [33] and [34]. It can also be noticed that the spectrum peaks occur at the same wavelength indicating a single NCS$^-$ ligand substitution by water [31], [35]. In reference to footnote 1, luminescence properties of Reinecke salt solution were also checked with our optical bench, no significant photon emission (either fluorescence or phosphorescence) at room temperature has been noticed in the 400-750 nm wavelength range in accordance with [36] (where the maximum luminescence quantum yield was estimated to be equal to $10^{-6}$ at 748 nm, figure that can be considered as negligible) or [29] and [37]. Moreover if photon would be emitted by the molecules at such a wavelength, they are weakly absorbed by the solution. As a consequence of these observations, no inner solution light source can be considered in applying the RTE in our reactors, all the photons that are present in the media are the ones emitted by the light sources.

At higher conversion according to [32], it appears that more than one NCS$^-$ ligand can be substituted by $H_2O$ molecule [38] leading to complex stoechiometry and reaction mechanisms and the compulsory determination of molar absorption coefficients. As a result of our experimental work on the subject (not presented in this publication), it is possible to state that this limits the confident use of Reinecke salt as an actinometer to a 1/3 molar conversion.

**3.2 Validation and analysis of our method with a small-scale and simple geometry reactor**

To validate our general methodology based on Monte Carlo Method for the RTE resolution as explained in the 2$^{nd}$ section, we first intend to carry out actinometrical experiments in a simple configuration (one dimensional, Cartesian geometry, normal quasi-collimated emission) using first a quasi-monochromatic and then a polychromatic light source. In such configurations, radiative transfer equation can be analytically solved (this point is developed in details in appendix B whereas the more general radiative transfer approach by Monte Carlo Method is presented in section 2). The calculated p values will be injected in the actinometer mass balance (Eq. 5); the differential equation will be numerically solved by Euler method for the estimation of $q_\cap$ by model inversion. Both approaches to describe the radiative transfer will be compared for the estimation of $q_\cap$.

### 3.2.1 Experimental set-up

The photoreactor developed for this 1D study is a flat and squared-section torus reactor intrapolated on the basis of a pilot plant used as photobioreactor [39]. This geometry presents two translucent faces made of glass: the first in front of the lighting source, the second at the rear of the reactor enabling the measurement of photon flux densities, $q_{\cap\ out}$, exiting of the reactor. In such a design the radiative transfer modelling can be correctly approximated as a one-dimension problem and, as already explained, analytically solved.

The torus geometry is obtained by the integration of cylindrical piece of metal in the centre of the reactor. Manufactured from a piece of 316L stainless steel, it can be seen in Fig. 3. The flow canal displays a 2.5 cm sided square section and the illuminated reactor surface, $S_{light}$, is equal to 59 cm². Depending on the filling volume, the reactor presents a dark fraction, $\varepsilon_d$, within a range of 7 to 10%; in such cases the specific illuminated surface, $a_{light}$, is around 37 $m^{-1}$.

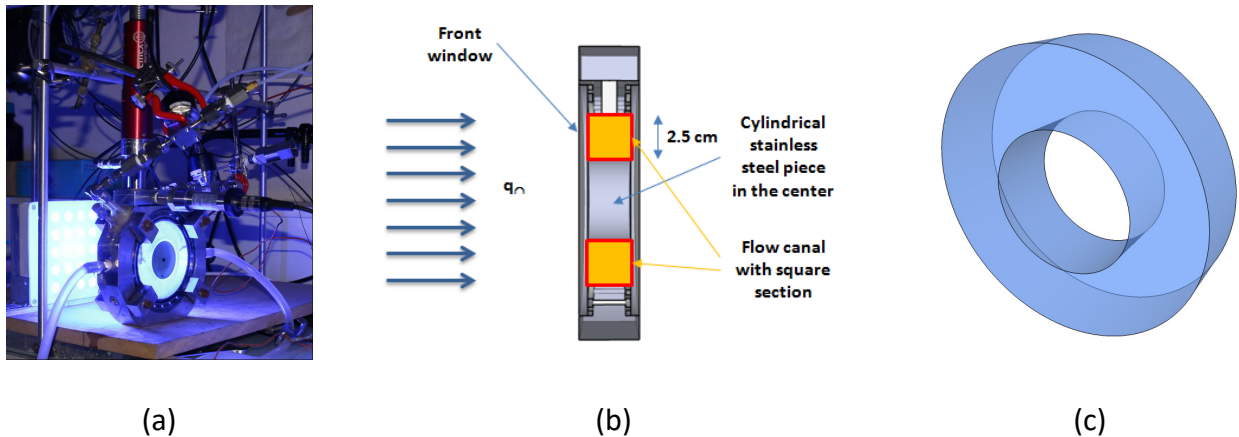

Fig. 3. (a) Photograph of the flat torus reactor and of the blue LED panel, (b) Horizontal cross section of the reactor presenting the square flow canal, (c) 3D CAD of the reactor simulated in the EDStar environment. Boundary conditions: reflectivity $\rho=0$, collimated normal incidence on the surface $S_{light}$ of the front window.

A liquid circulation circuit is machined in the stainless steel piece, in parallel to the flow canal. It enables the flow of a coolant fluid (water for example) using thermostatic bath/circulator (Lauda eco RE 415); a custom built RTD Pt100 sensor (TCDirect) is laterally positioned in order to set the reactor temperature at 25°C. A lid is located at the top of the reactor; it offers additional possibilities for inlets or outlets and a support for the mixing device. It is constituted of a micromotor 24V/DC (Minisprint Magnetic Stirrer, Premex Reactor ag) using magnetic coupling technology that ensures gas tightness. The motor rotation speed is controlled with a digital speed display DZA-612Z. Connected to the motor by a shaft, an impeller makes the liquid rotate in the annular space, mixing the solution. Additional information can be found in [40].

The light sources utilized with the torus reactor are LED panels fabricated by Sibylux. The first one is composed of 25 (5×5) LED (Royal blue D42180, Seoul Semiconductor) equipped with lenses. It provides a blue quasi-collimated light on a 12.5×12.5 cm surface. Such a monochromatic light source has been adopted for ease of use and post treatment. The second panel includes 64 (8×8) white LED also equipped with lenses; it generates a collimated light on a 16×16 cm surface. The emission spectra and the emission probability density function, $p_\Lambda^{light}(\lambda)$, of the two panels were determined with a USB 2000+ Ocean Optics spectrometer (see Fig. 4). The blue LED provides a blue light with a Gaussian shaped emission spectrum with a maximum at a 457 nm wavelength. The emission spectrum of the white LED indicates it is composed of a blue LED and a yellow phosphor, these two complementary colours combine to form white light [41].

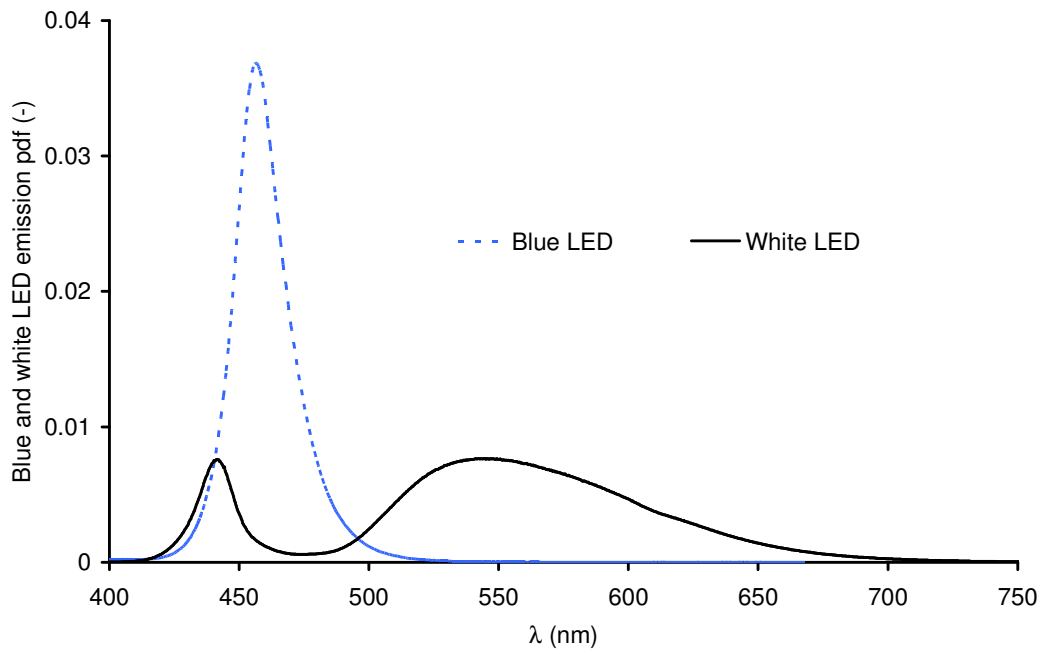

Fig. 4. Emission spectra $p_\Lambda^{light}(\lambda)$ of the LED utilized with the torus reactor measured with a USB 2000+ Ocean Optics spectrometer.

For both panels, the photon flux density can be easily and accurately controlled via an USB DMX controller and the Easy Stand Alone software (Nicolaudie), including 256 different setting positions to modify the electric power supplied to the LEDs and thus the emitted photon flux density, $q_\cap$, whose value will be determined with software settings.

Previous residence time distribution experiments (presented in another publication [40]) were carried out to ascertain the hydrodynamic behaviour of the torus reactor. Under the considered experimental conditions it can be described as a perfectly stirred reactor, the actinometer concentration will therefore be considered as homogeneous throughout in the reactor. Its evolution could be modelled with Eq. 5 where the only unknown is the photon flux density (to be determined by model inversion).

**3.2.2 Validation of the Monte Carlo algorithm for radiative transfer analysis**

For the further use of the complete model associated with Monte Carlo algorithm in complex geometries, it is possible to validate here in a simple 1D experimental geometry (where RTE could be analytically written) its implementation .

Hence the validation of our Monte Carlo Method for radiative transfer can be made through the comparison of its predicted concentrations, the predicted concentrations by the analytical radiative transfer model and experimental kinetics of the actinometer decomposition. On Fig. 5 is represented an example of thiocyanate concentration evolution against time (data with • symbol) for a given setting of the LED panel control software (*i.e.* an unknown incident photon flux density $q_\cap$). This unidentified value was determined by comparing experimental results and the mass balance models (Eq. 5) using the two different radiative transfer analyses. As already explained the first approach consists in the implementation of a Monte Carlo Method (to solve the radiative transfer equation in this experimental configuration (see section 2) modelled a by three-dimensional CAD (see Fig. 3.c)). The second method consists in using Eq B.3 a simple and analytical expression for estimation of p.

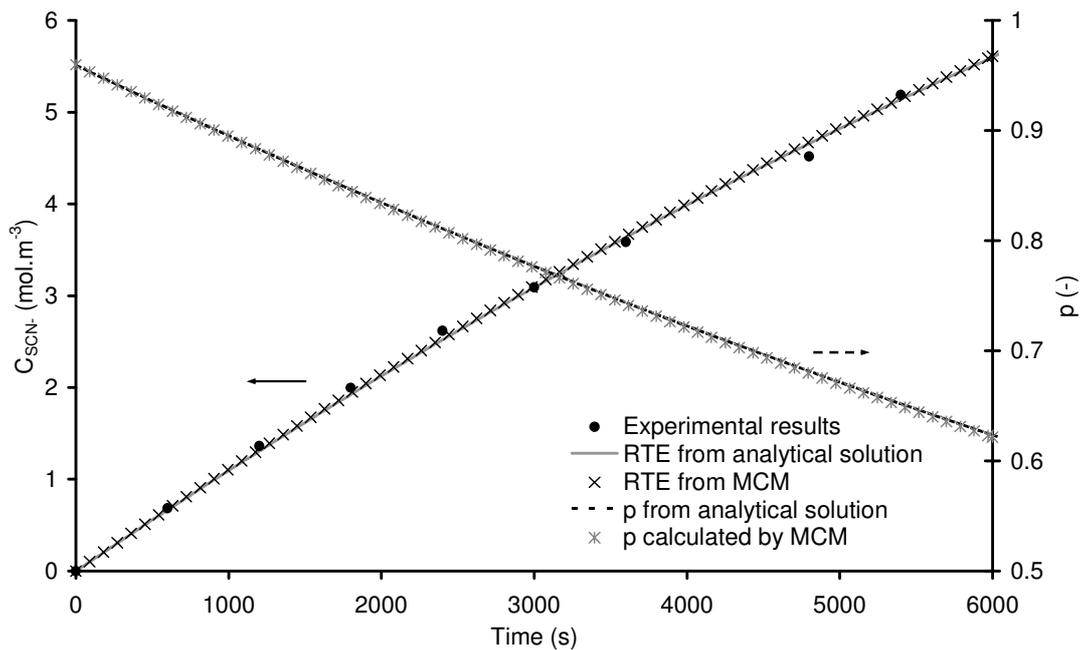

Fig. 5. Example of experimental results obtained in the flat torus reactor (● symbols), comparison between analytical method (lines) and Monte Carlo Method (cross symbols) for kinetics (left ordinate axis) and absorption probability p (right ordinate axis) for $q_\cap$ = 116 $\mu mol_{h\nu}.m^{-2}.s^{-1}$.

As mentioned earlier, we chose to restrict the photon flux density determination to a limited concentration range in thiocyanate, meaning a low conversion (around 1/3); nonetheless this corresponds to an analysis beyond the usual linear zone.

Hence the adequacy of the radiative models with the experimental data is checked. A photon flux density can then be estimated, in the case of Fig. 5 experimental results $q_\cap$ value is equal to 116 µmol.m$^{-2}$.s$^{-1}$ in both cases. The two methods used to model radiative transfer (Monte Carlo Method or the analytic expression of p (Eq. B.3)) injected in the mass balance succeed in correctly describing the evolution of thiocyanate concentration. Our approach for solving radiative transfer using MCM in the case of 1D experimental actinometry can be considered as validated.

The probability p that a photon entering the system is absorbed by the actinometer is also presented in Fig. 5. At the beginning of the experiment, p is inferior to 1, a sign that the photon absorption is incomplete in our flat torus reactor considering the absorption optical thickness e is roughly equal to 2.9 (=7.8×15×0.025), is not high enough. This point is in agreement with photon flux density measurement at the back of the reactor during experiment. As the reaction proceeds and as Reinecke salt is transformed, p is decreasing. As expected (the predicted kinetics are identical), the p probabilities (estimated by MCM and the analytical method (Eq. B.3) are superimposing in Fig. 5.

In order to fully check the validity of our methodology (photon flux identification using kinetics), the hemispherical incident photon flux densities (PFD), $q_\cap$ in the region of interest (the canal for liquid) was also determined via a physical method using a LiCOR quantum sensor (Li-190Sa) connected to a Li-189 display and by averaging 16 points of measurement on a plane parallel to the LED panel (representing 15 % of the total illuminated surface).

This typical procedure ([10], [42]) using a physical method permits the determination of the mean $q_\cap$ value and of its standard deviation. For both LED panels, mean $q_\cap$ increases linearly with software settings (not shown here). Considering the standard deviation, in the case of blue LED, it is equal to 11%, mainly due to edge effect (*i.e.* surface inhomogeneity), the LED panel having only 5 LEDs in height and width. A 3.5% standard deviation is reached with the white LED panel thanks to the higher LED number.

For blue and white LED panels, the mean $q_\cap$ values determined by quantum sensor are respectively represented in the abscissa axis of Fig. 6 a and b which show the parity diagram between photon flux determined by the quantum sensor and the mean (obtained from 3 different actinometry experiments) identified photon flux density value using three degrees of refinement for the models. Developed for collimated emission sources, these models are respectively called spectral model with by-product, *i.e.* the more rigorous model, grey model with by-product and a grey model without by-product as developed by ([10]) after the phenomenon they take into account.

As can be seen in Fig. 6.a, the use of grey model without by-product results in an underestimation of photon flux density (with a slope of 0.72±0.05) compared to quantum sensor measurements. This can be explained by the quasi-equality of the absorption coefficient $E_i$ for Reinecke salt and by-product. The by-product cannot be ignored; the related model can no longer be used for a correct and accurate estimation of photon flux density even in 1D geometry. On the other hand the results are very satisfying for the last two models (spectral and grey model with by-product); the identified flux densities are close to the line of perfect agreement. The slopes are respectively equal to 1.00±0.04 and 0.97±0.04. Due to the narrow emission spectrum of the blue LED (see Fig. 4), the results obtained using grey model with by–product are nearly equal to that of the spectral model. Actinometry experiment using Reinecke salt should take into consideration the presence of the by-product for the accurate estimation of photon flux densities.

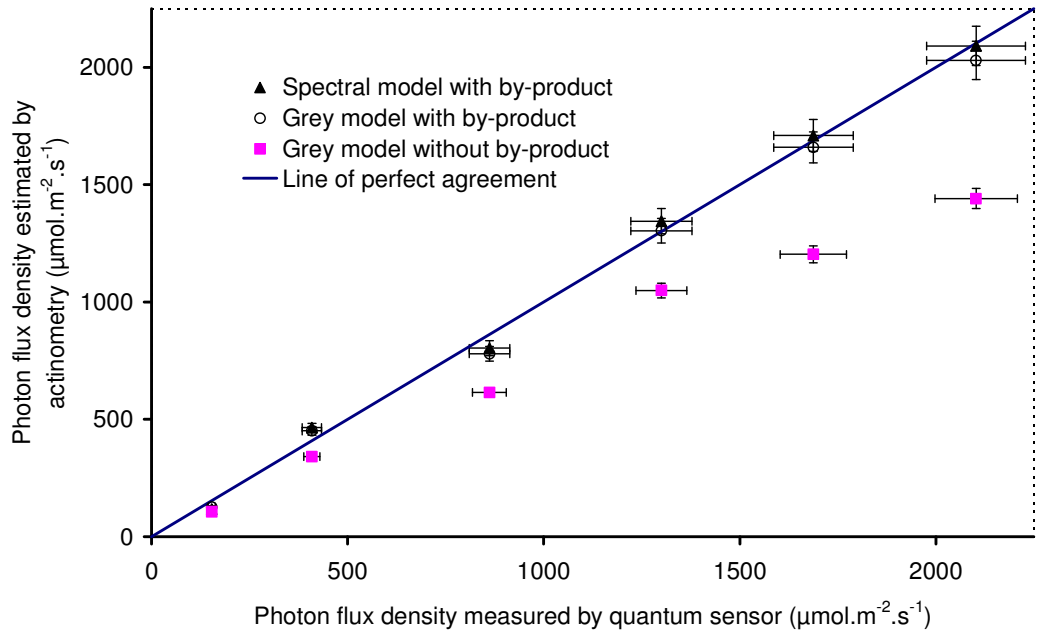
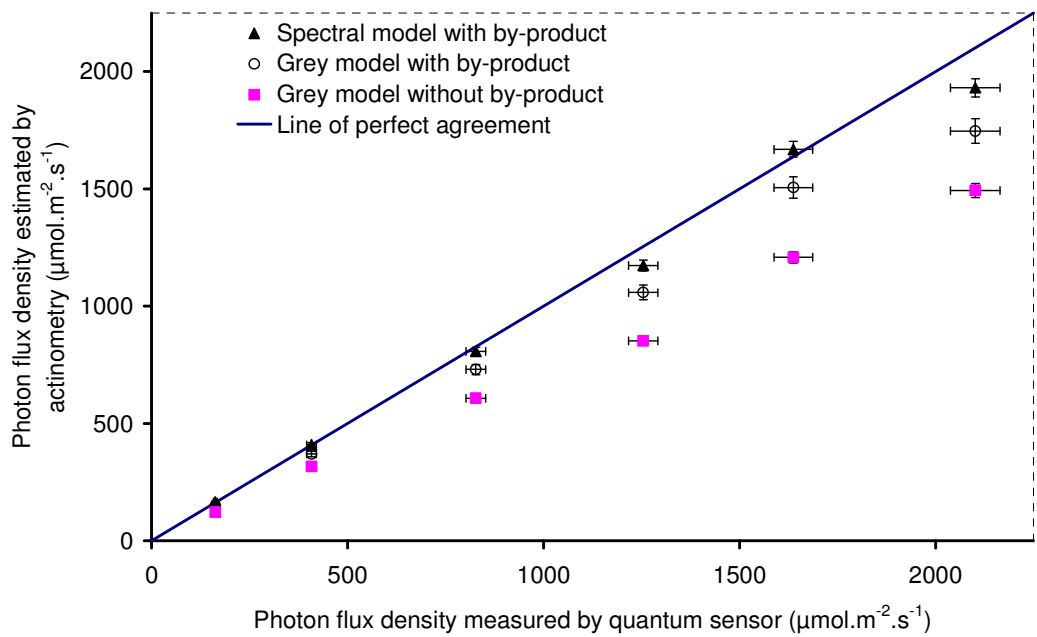

Fig. 6 Parity diagrams for the determination of the photon flux density (actinometry/quantum sensor) obtained in the flat torus reactor using blue LED (a) and white LED (b) panels.

In the case of a real polychromatic light source, these results are different. Indeed as can be seen in Fig. 6(b), the spectral model with by-product follows the parity line, whereas the grey model with by-product differs from it (the slopes are respectively equal to 0.96±0.05 and 0.86±0.05). It is thus compulsory to take into account the spectral aspect of the light source.

For additional information, the identified photon flux densities for grey model without by-product are also plotted, confirming the already noticed discrepancy of this approach (with a 0.72±0.03 slope).

In this section, we thus succeed in validating the actinometry method and our related result treatment methodology, based on the estimation of probability absorption p, in a simple experimental geometry (1D Cartesian) with quasi-collimated light source by comparing identified photon flux density values with quantum sensor measurements. The use of a physical method (sensor) or control sample containing the actinometer [43] is no longer possible in complex geometry making our radiative transfer methodology essential for the estimation of photon flux density.

### 3.3 Implementation in a real-world pilot-plant photobioreactor with complex geometry

#### 3.3.1 Presentation of the experimental system and its use in actinometry

As a result of our previous work [12], we succeeded in determining the optimum design for volumetrically lightened photobioreactors using knowledge models and the constructal approach. This resulted in the concept of DiCoFluV reactor (an acronym for "dilution controlée du flux en volume" [12]); its functioning needs to be characterized with artificial illumination, before being tested in true solar conditions.

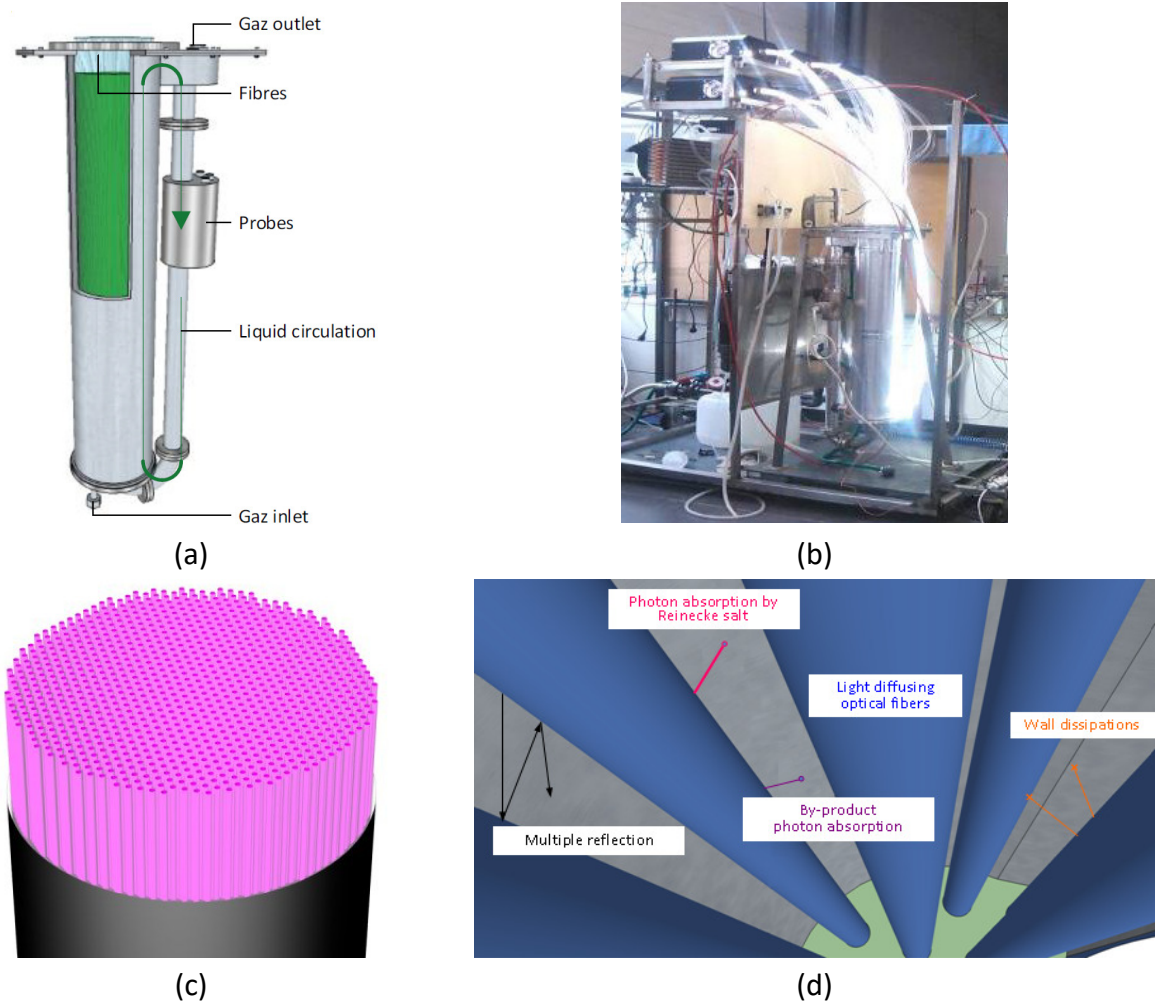

Fig. 7. (a) Scheme of the air-lift reactor DiCoFluV (b) Photograph of the complete installation (including the reactor, lamps and optical fibers), c and d) CAD of the reactor simulated in the EDStar environment: (c) optical fibres layout within the cylindrical reactor tank (d) representation of the phenomena simulated in the complex geometry of the reactor. Boundary conditions: reflectivity of the stainless steel reactor tank is 0.54, reflectivity of the optical fibres is 0.1, diffuse reflection $p_\Omega^\rho(\boldsymbol{\omega}, \mathbf{x}) = \dfrac{\boldsymbol{\omega} \cdot \mathbf{n}}{\pi}$ where **n** is the inner normal at location **x**, both collimated normal and diffuse (i.e. lambertian distribution $p_{\Omega_0}^{light}(\boldsymbol{\omega_0}, \mathbf{x}) = \dfrac{\boldsymbol{\omega_0} \cdot \mathbf{n}}{\pi}$) emission at the surface $S_{light}$ of the fibres are investigated.

The conception of the DiCoFluV photoreactor is based on an airlift system (Fig 7.a). The reactor has two attached parts: the riser with an internal diameter of 165 mm and the downcomer tube of 50 mm internal diameter. The whole vessel is made of 316L stainless steel. The temperature is controlled thanks to a water jacket around the riser. The gas inlet at the bottom of the riser creates a turbulent liquid circulation that mixes the reaction media. The gas is injected through an annular porous stainless steel plate around the downcomer return tube (see further) at a flow rate of 21 L.min$^{-1}$, corresponding to 1 vvm. With this flow rate the mixing time is about 30 seconds. The gas is separated from the liquid phase at the top of the reactor: the liquid phase goes in the downcomer tube and the gas through the gas outlet. The gas goes through a condenser that drives the condensed water back to the reactor. Temperature and pH sensors are positioned in the downcomer tube. The dark fraction, $\varepsilon_d$, is measured to be 10%.

The light is provided by six discharge metal halide lamps (BLV, 270 411 MHR 250N) through 977 PMMA optical fibres immersed in the riser (see Fig. 9 for the emission spectrum at the optical fibres). The illuminated surface, $S_{light}$, is thus estimated to 7.4 m$^2$ and the specific illuminated surface, $a_{light}$, is equal to 370 m$^2$.m$^{-3}$ far higher than the one of torus photoreactor presented in section 3. The fibres have been laser impacted on their immersed surface to diffuse the light laterally. They are not directly in contact with the liquid (that could be corrosive): they are thread into polycarbonate tubes (of 2.4 mm external diameter). Those tubes are attached to a 316L stainless steel lid with epoxy glue in a hexagonal lattice (centre to centre distance: 4.8 mm).

As can be seen in Fig. 7.b, the geometry can be considered as complex (with possible interactions between photons and polycarbonate tubes or reactor wall), the photon absorption by actinometer as incomplete (the order of magnitude of the optical thickness, e, is 0.5) and, of course, the light source as polychromatic. Moreover photon emission model at the fibers *i.e.* the angular distribution is unknown unlike in the flat torus system (where it is known as quasi-collimated by construction). The two limit cases (lambertian and collimated) will be addressed. In face of all this complexity, the methodology for radiative transfer description developed in section 2 and shown in Fig. 8 makes sense. Hence the algorithm that describes photon transport will be the same as the one utilized and validated in section 3. Only the geometry, managed by the EDStar development environment, will be different.

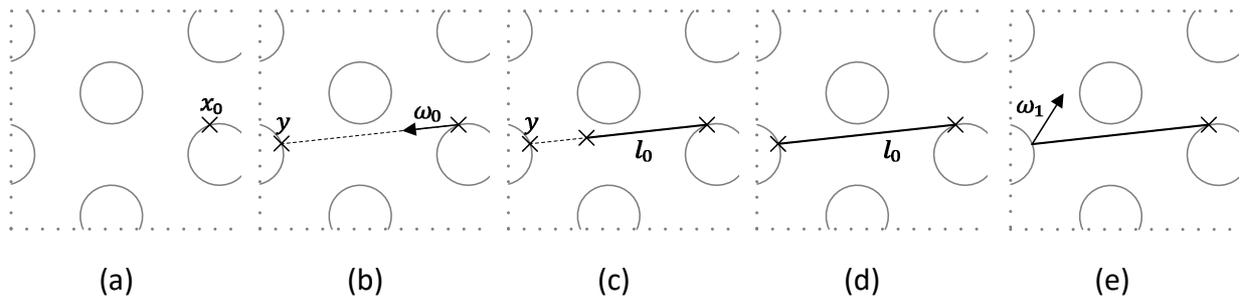

(a)  (b)  (c)  (d)  (e)

Fig. 8. Scheme of the optical-path sampling procedure in the case of the DiCoFluV reactor

(a) Uniform sampling of an emission location $x_0$, (b) Sampling of the emission direction $\omega_0$, (c) Sampling of the absorption length, $l_0$, and determination of the intersection **y** with bonding surface S. In that case $l_0$ is shorter than $||y - x_0||$, the photon is absorbed (d) $l_0$ is longer than $||y - x_0||$, the optical path interacts with the surface S , where the photon is absorbed (e) $l_0$ is longer than $||y - x_0||$, the optical path interacts with the surface S where the photon is reflected with a direction $\omega_1$.

Concerning the implementation of the actinometric reaction, due to large volume to be used, the Reinecke salt solution is directly prepared in the reactor, with the lamps turned off. Procedure explained in [10] is once again followed. Hence the salt is dissolved at 40°C in about 21 L of a $10^{-3}$ mol.L$^{-1}$ potassium hydroxide solution (at pH=11). As the dissolution of the salt makes the pH drop, 10 to 14 mL of a commercial 17.6 M potassium hydroxide solution are added in order to maintain the pH around a value of 11. The pH stays stable when the salt is totally dissolved: the temperature is then lowered to 23°C. The stripping of the ammoniac is easily made with the air-lift system. Theoretically (considering a $k_L a$ of 150 h$^{-1}$ and a gas flow rate of 20 L.min$^{-1}$) the stripping should be over in less than 10 minutes. The pH is then lowered between 3 and 5 by adding a concentrated (97%) commercial sulphuric acid solution. About 500 mL of the solution obtained are extracted from the reactor and kept away from light to be used as the reference solution.

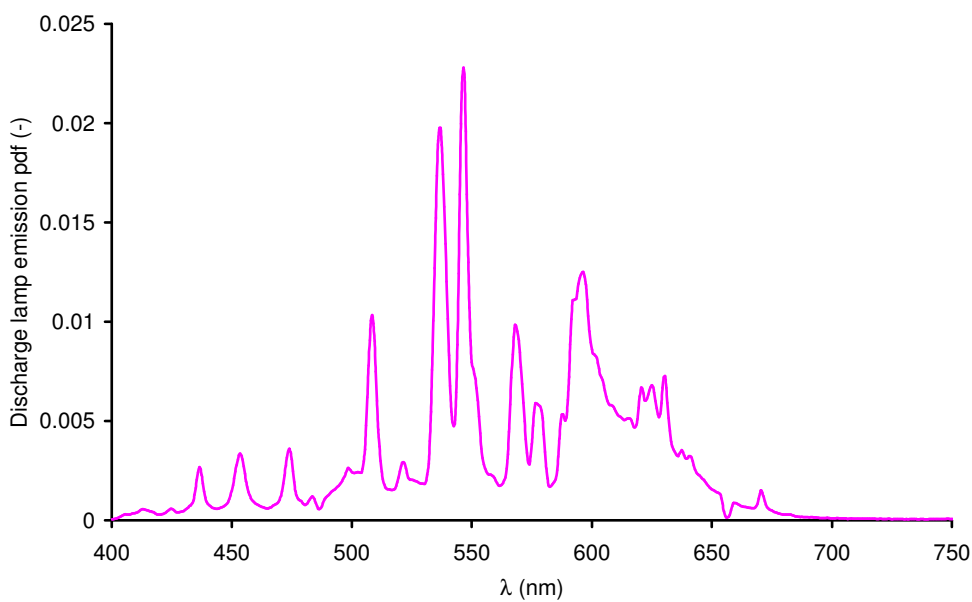

Fig. 9. Emission spectrum of the discharge lamps used in the DiCoFluV measured at the surface of optical fibers with a USB 2000+ Ocean Optics spectrometer.

### 3.3.2. Actinometry results in complex geometry

Four actinometry experiments have been carried out in the DiCoFluV reactor, all beyond linearity range; the measured thiocyanate ion evolutions against time for each experiment are presented in Fig. 10. Given the small dispersion of concentrations obtained in such complex pilot plant, reproducibility is ensured in our experiments.

As already explained, these results are essential for the determination of the photon flux density and the angular distribution of photon emission, the two degrees of freedom of our inversion problem.

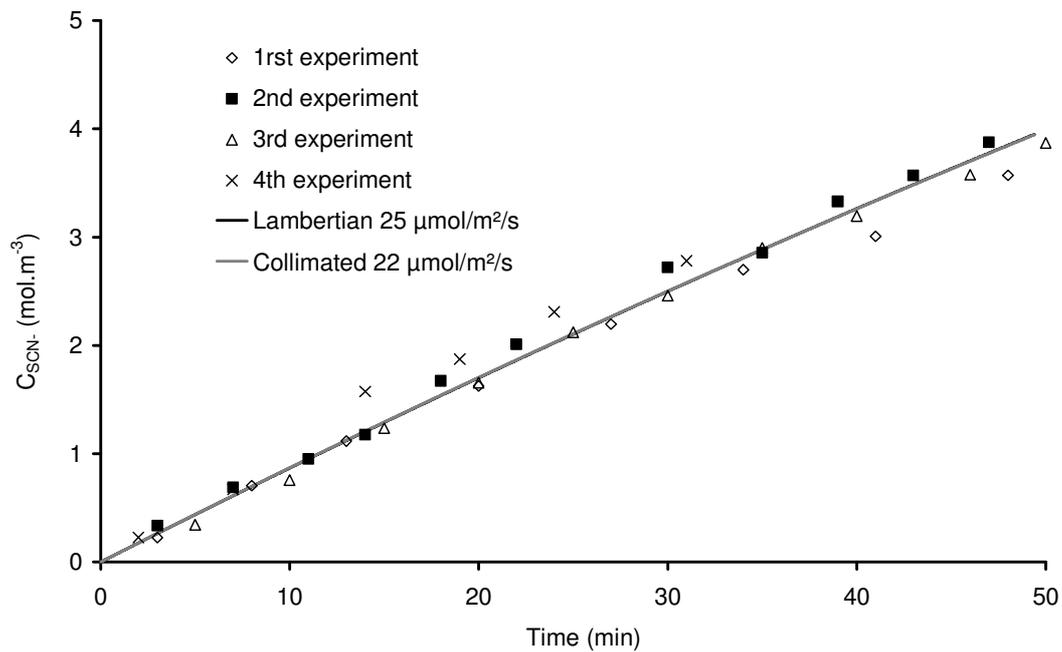

Fig. 10. Kinetic results obtained in the DiCoFluV reactor ($C_0$ = 15 mol.m$^{-3}$) and predicted SCN$^-$ evolution by the two emission models (lambertian and collimated) as a result of the inversion procedure for determination of $q_\cap$.

Based on the results obtained in simple mono-dimensional geometry (see section 3), grey models with mean absorption coefficients couldn't be used to accurately estimate photon flux density emitted by the discharge lamp. Only the polychromatic aspect of the light source is considered in our models. In addition, the two emission limit cases are considered: lambertian and collimated; the photon flux densities are identified for these two configurations. An example of photon flux density estimation by the two models is presented in Fig. 10, for all the experimental data. In that case, it is possible to notice a good fit of experimental data by both models. Identified photon flux densities with extended uncertainties values are respectively equal to $(22 \pm 2)$ µmol.m$^{-2}$.s$^{-1}$ for lambertian emission and $(25 \pm 2)$ µmol.m$^{-2}$.s$^{-1}$ for collimated emission. Considering experimental error, the match between models and experiment appears very good. No difference could be made in this range of conversion between the predicted concentrations by the two models and their respective angular distribution as boundary condition. If the classical linear treatment of actinometry was used, neglecting radiative transfer, a $(15 \pm 1)$ µmol.m$^{-2}$.s$^{-1}$ photon flux density would be identified, leading to an unsatisfactory 50 percent error.

## 4. Conclusion and perspectives

### 4.1 New perspectives in using actinometry as a mean to obtain angular information on the light source emission in any geometry

As demonstrated in previous section, the direct and simultaneous determinations of angular emission model and photon flux density aren't possible. Hence to succeed in estimating those two physical values, the following strategy can be used (provided a hypothetical actinometer allowing high conversion without complication is developed):

- working at high actinometer concentration. This will enable to determine $q_\cap$ whatever the angular emission model when working at low conversion (< 0.5). To illustrate this comment, using the optical properties of Reinecke's salt system, thiocyanate ion concentration evolutions were calculated using Eq. 5 and Eq. B.3 for collimated source and Eq. 5 and Eq B.5 for lambertian source for a 1D configuration. The results can be seen in Fig. 11.a), where the predicted dimensionless thiocyanate concentrations (for a 15 mol.m$^{-3}$ initial actinometer concentration, corresponding to typical experiment conditions used in the present article) are plotted as function of dimensionless time, $t/\tau$, for the two emission models at a given photon flux density. The concentration evolutions predicted by the two models cannot be distinguished before high conversion due to the high optical thickness. Photon flux density, $q_\cap$, can be determined independently of the emission model.

- operating with diluted solutions will permit to easily discriminate between the two limit cases of emission knowing the photon flux density. In the same way as in previous paragraph, thiocyanate ion concentration evolutions were calculated for a 0.15 mol.m$^{-3}$ actinometer concentration. As can be seen in Fig. 11.b where the predicted dimensionless thiocyanate concentrations are plotted even at low conversion, the differences in dimensionless thiocyanate concentrations are evident.

Hence a new challenge for the future of actinometry would be the development of a new actinometer with the following properties:

- a wide usable spectral range in the visible wavelength, associated with sufficiently high molar absorption coefficient, E,

- simple photoreaction mechanism (only 1 by-product that could absorb or not photon) and preparation protocole, in addition with easy analytical methods for titration,

- achievement of high conversion (ferrioxalate isn't suitable as precipitation occurs at even low conversion),

- low cost for an extensive use in large scale photo-processes.

Another aspect of the development of actinometry in complex geometry would be to model the emission source. This point has already been presented elsewhere [43] [44] [45] in simple configurations, however such a procedure can not be used in our systems, where emission is more complicated than fluorescent tubes. It is thus compulsory to model photon transport in the optical fibres from the lamp till their exit the reactor, using our Monte Carlo methodology in association with the EDStar library. This scientific aspect is under investigation by our group.

## 4.2 Conclusions

In this article, a complete radiative transfer approach for the estimation of photon flux density by actinometry has been presented. First of all to help the reader, the traditional and classical treatment of actinometry is explained with the associated restrictions. It can no longer be used in our research area aimed at studying natural or artificial photosynthesis in intensified processes with low optical thickness, that's why a radiative transfer analysis, resolved by the latest advances in Monte Carlo Method, is used to describe actinometry in simple and in complex geometries.

The chosen actinometer, the Reinecke salt, is thoroughly characterized: its optical properties and the ones of the by-product have been precisely determined and documented in an appendix available for any interested user.

A second step has been to use the actinometer in a simple monodimensional photoreactor and to irradiate it. This has been successively done with monochromatic and polychromatic LED panels, whose mean hemispherical photon flux densities emitted have been measured by hemispherical cosine sensor mapping. Under irradiation, the Reinecke salt decomposition has been monitored and photon flux densities have been estimated based on different developed models. The comparison with the mapping has given excellent matches especially for the complete model taking into account polychromatic light and by-product absorption.

After validating our radiative transfer approach by MCM in simple geometry, the Reinecke salt has been for the first time used in a 25 L reactor presenting a complex geometry, the DiCoFluv photoreactor, where quantum sensors are unusable and the optical thickness is low. Despite these difficulties, our Monte Carlo Methodology, in connection with the EDStar development environment in charge of managing the complex geometry, has enabled us to easily estimate the mean photon flux density. The only remaining unknown in our system is the angular emission model of the light source. To solve this problem, potential sources of improvement have been laid out especially on the requirements of a new actinometer and modelling of the photon emission.

**Acknowledgment**

This work has been sponsored by the French government research-program "Investissements d'avenir" through the ANR programs BIOSOLIS (2008-11), Tech'Biophyp (2011-2015), PRIAM (2013-15). This work has also been sponsored through the IMobS3 Laboratory of Excellence (ANR-10-LABX-16-01), by the European Union through the Regional Competitiveness and Employment program -2007-2013- (ERDF – Auvergne region) and by the Auvergne region. It is also founded by the CNRS through the PIE program PHOTORAD (2010–11) and the PEPS program "Intensification des transferts radiatifs pour le développement de photobioréacteurs à haute productivité volumique" (2012–13). The authors acknowledge the CNRS research federation FedESol where fruitful discussions and debates take place every year since 2012. The authors also acknowledge Méso-Star SAS for fruitful discussions about ray tracing in complex geometry and orthogonal programming (www.meso-star.com). Finally, the authors are grateful to MM. Pascal Lafon and Frederic Joyard for their invaluable assistance regarding the developments of the torus and DiCoFluV photoreactors.

## Appendix A: Spectral absorption coefficients of the Reinecke salt and the monosubstituted chromium complex

| λ (nm) | $E_a$ Reinecke salt ($m^2 \cdot mol^{-1}$) | $E_a$ [Cr(NCS)$_3$H$_2$O(NH$_3$)$_2$]$^-$ ($m^2 \cdot mol^{-1}$) | λ (nm) | $E_a$ Reinecke salt ($m^2 \cdot mol^{-1}$) | $E_a$ [Cr(NCS)$_3$H$_2$O(NH$_3$)$_2$]$^-$ ($m^2 \cdot mol^{-1}$) |
|---|---|---|---|---|---|
| 400 | 21.73 | 15.95 | 576 | 12.45 | 10.41 |
| 402 | 21.33 | 15.64 | 578 | 11.96 | 9.94 |
| 404 | 20.90 | 15.28 | 580 | 11.45 | 9.50 |
| 406 | 20.35 | 14.90 | 582 | 10.89 | 9.12 |
| 408 | 19.63 | 14.49 | 584 | 10.36 | 8.72 |
| 410 | 18.71 | 14.08 | 586 | 9.84 | 8.34 |
| 412 | 17.54 | 13.58 | 588 | 9.29 | 8.02 |
| 414 | 16.70 | 13.07 | 590 | 8.78 | 7.70 |
| 416 | 15.54 | 12.75 | 592 | 8.27 | 7.39 |
| 418 | 14.60 | 12.11 | 594 | 7.84 | 7.05 |
| 420 | 13.63 | 11.48 | 596 | 7.42 | 6.73 |
| 422 | 12.66 | 10.83 | 598 | 6.99 | 6.38 |
| 424 | 11.68 | 10.24 | 600 | 6.68 | 6.03 |
| 426 | 10.70 | 9.70 | 602 | 6.12 | 5.75 |
| 428 | 9.73 | 9.21 | 604 | 5.73 | 5.45 |
| 430 | 8.85 | 8.78 | 606 | 5.34 | 5.15 |
| 432 | 8.29 | 8.42 | 608 | 4.99 | 4.91 |
| 434 | 7.93 | 8.08 | 610 | 4.63 | 4.69 |
| 436 | 7.52 | 7.78 | 612 | 4.25 | 4.52 |
| 438 | 7.22 | 7.51 | 614 | 3.95 | 4.31 |
| 440 | 7.04 | 7.27 | 616 | 3.67 | 4.10 |
| 442 | 6.90 | 7.07 | 618 | 3.36 | 3.85 |
| 444 | 6.85 | 6.92 | 620 | 3.09 | 3.58 |
| 446 | 6.84 | 6.83 | 622 | 2.86 | 3.36 |
| 448 | 6.92 | 6.78 | 624 | 2.61 | 3.15 |
| 450 | 7.01 | 6.79 | 626 | 2.39 | 2.95 |
| 452 | 7.18 | 6.85 | 628 | 2.15 | 2.77 |
| 454 | 7.36 | 6.97 | 630 | 1.98 | 2.60 |
| 456 | 7.66 | 7.15 | 632 | 1.82 | 2.44 |
| 458 | 8.11 | 7.41 | 634 | 1.67 | 2.28 |
| 460 | 8.63 | 7.71 | 636 | 1.54 | 2.14 |
| 462 | 9.12 | 8.07 | 638 | 1.37 | 2.01 |
| 464 | 9.70 | 8.49 | 640 | 1.25 | 1.88 |
| 466 | 10.24 | 8.93 | 642 | 1.14 | 1.77 |
| 468 | 10.85 | 9.39 | 644 | 1.04 | 1.66 |
| 470 | 11.62 | 9.86 | 646 | 0.92 | 1.55 |
| 472 | 12.32 | 10.35 | 648 | 0.81 | 1.46 |
| 474 | 13.06 | 10.83 | 650 | 0.71 | 1.37 |
| 476 | 13.91 | 11.33 | 652 | 0.65 | 1.28 |
| 478 | 14.58 | 11.82 | 654 | 0.58 | 1.20 |
| 480 | 15.34 | 12.35 | 656 | 0.49 | 1.13 |
| 482 | 16.09 | 12.90 | 658 | 0.44 | 1.06 |
| 484 | 16.75 | 13.36 | 660 | 0.41 | 0.99 |
| 486 | 17.42 | 13.90 | 662 | 0.43 | 0.93 |
| 488 | 18.08 | 14.43 | 664 | 0.43 | 0.87 |
| 490 | 18.78 | 14.90 | 666 | 0.41 | 0.82 |
| 492 | 19.41 | 15.39 | 668 | 0.38 | 0.77 |
| 494 | 20.07 | 15.82 | 670 | 0.33 | 0.72 |
| 496 | 20.84 | 16.17 | 672 | 0.28 | 0.67 |
| 498 | 21.49 | 16.55 | 674 | 0.22 | 0.63 |
| 500 | 21.99 | 16.86 | 676 | 0.17 | 0.59 |
| 502 | 22.51 | 17.25 | 678 | 0.13 | 0.56 |
| 504 | 22.90 | 17.57 | 680 | 0.09 | 0.52 |
| 506 | 23.28 | 17.96 | 682 | 0.06 | 0.49 |
| 508 | 23.61 | 18.20 | 684 | 0.04 | 0.46 |
| 510 | 23.92 | 18.40 | 686 | 0.02 | 0.43 |
| 512 | 24.14 | 18.57 | 688 | 0.01 | 0.40 |
| 514 | 24.31 | 18.71 | 690 | 0.00 | 0.38 |
| 516 | 24.45 | 18.81 | 692 | 0.00 | 0.35 |
| 518 | 24.53 | 18.88 | 694 | 0.00 | 0.33 |
| 520 | 24.52 | 18.92 | 696 | 0.01 | 0.31 |
| 522 | 24.56 | 18.92 | 698 | 0.01 | 0.29 |
| 524 | 24.51 | 18.90 | 700 | 0.02 | 0.27 |
| 526 | 24.40 | 18.84 | 702 | 0.03 | 0.26 |
| 528 | 24.20 | 18.75 | 704 | 0.04 | 0.24 |
| 530 | 23.87 | 18.64 | 706 | 0.05 | 0.23 |
| 532 | 23.55 | 18.49 | 708 | 0.06 | 0.21 |
| 534 | 23.13 | 18.32 | 710 | 0.07 | 0.20 |
| 536 | 22.69 | 18.12 | 712 | 0.08 | 0.19 |
| 538 | 22.26 | 17.90 | 714 | 0.08 | 0.17 |
| 540 | 21.78 | 17.65 | 716 | 0.09 | 0.16 |
| 542 | 21.25 | 17.44 | 718 | 0.09 | 0.15 |
| 544 | 20.96 | 17.08 | 720 | 0.09 | 0.14 |
| 546 | 20.31 | 16.71 | 722 | 0.09 | 0.14 |
| 548 | 19.82 | 16.33 | 724 | 0.09 | 0.13 |
| 550 | 19.27 | 15.96 | 726 | 0.08 | 0.12 |
| 552 | 18.77 | 15.55 | 728 | 0.08 | 0.11 |
| 554 | 18.30 | 15.12 | 730 | 0.08 | 0.10 |
| 556 | 17.75 | 14.73 | 732 | 0.07 | 0.10 |
| 558 | 17.24 | 14.27 | 734 | 0.06 | 0.09 |
| 560 | 16.70 | 13.84 | 736 | 0.06 | 0.09 |
| 562 | 16.21 | 13.38 | 738 | 0.05 | 0.08 |
| 564 | 15.72 | 12.91 | 740 | 0.05 | 0.08 |
| 566 | 15.21 | 12.43 | 742 | 0.04 | 0.07 |
| 568 | 14.63 | 12.05 | 744 | 0.04 | 0.07 |
| 570 | 14.10 | 11.64 | 746 | 0.04 | 0.06 |
| 572 | 13.59 | 11.19 | 748 | 0.04 | 0.06 |
| 574 | 13.04 | 10.78 | 750 | 0.03 | 0.05 |

Previous scientific approaches (for example [10]) didn't take completely into account the spectral emission of light source $\rho_\Lambda^{light}(\lambda)$ (as can be seen in Fig. 4) and the spectral distribution of absorption coefficients of reagent and product (Fig. 2). Instead, mean absorption coefficient could be estimated using the value at maximum lamp emission wavelength or (in a more correct way) utilizing the emission probability density function of the light source:

$$E_{i,mean} = \frac{\int_{\lambda_{min}}^{\lambda_{max}} E_{i,\lambda}\, \rho_\Lambda^{light}(\lambda)\, d\lambda}{\int_{\lambda_{min}}^{\lambda_{max}} \rho_\Lambda^{light}(\lambda)\, d\lambda} \quad (A.1)$$

Finally, this method is still approached as it assumes equal photon distribution with respect to the wavelength. Models developed on such average will be considered as grey.

**Appendix B: Simple analytical solution of the radiative transfer equation in one-dimensional configuration with non-reflecting surfaces and monochromatic light-source**

Hereafter we detail how the analytical solutions for p Eq. 10 simplifies for one-dimensional configuration with non-reflecting surfaces and monochromatic radiation.

The reaction volume is a slab contained within abscissa x=0 and x=L, with incident radiation at x=0. For this one-dimensional configuration, the integral $\int_S dx_0 \frac{1}{S}$ in Eq. 10 vanishes and the condition $x_1 \in V$ writes $0 < x_1 < L$. For monochromatic sources, the spectrum $p_\Lambda^{light}(\lambda)$ is a delta-Dirac distribution centered at wavelength $\lambda_i$ of emission; therefore the integral $\int_{\lambda_{min}}^{\lambda_{max}} d\lambda \, p_\Lambda^{light}(\lambda)$ in Eq. 10 vanishes and every wavelength dependent parameters are taken at $\lambda=\lambda_i$. Finally, for non-reflecting surfaces $\rho=0$ the integral formulation is no more recursive and overall, Eq. 10 becomes

$$p = \int_{2\pi^+} d\omega_0 \, p_{\Omega_0}^{light}(\omega_0) \int_0^{+\infty} dl_0 \, k_{\lambda_i} \exp(-k_{\lambda_i} l_0) H(0 > x_1 > L) \frac{C_A E_{A,\lambda_i}}{k_\lambda} \tag{B.1}$$

where the Heaviside notation $H(0 > x_1 > L)$ takes the value 1 if $0 < x_1 < L$ and 0 otherwise. For the one-dimensional configuration addressed here, with incident radiation at x=0, we have $x_1 = \mu_0 \cdot l_0$, where $\mu_0 = \omega_0 \cdot e_x$ is the cosine between the incident direction $\omega_0$ and the unit vector $e_x$ along the x axis. Therefore $\int_0^{+\infty} dl_0 \, k_{\lambda_i} \exp(-k_{\lambda_i} l_0) H(0 > x_1 > L)$ in Eq. B.1 becomes

$$\int_0^{+\infty} dl_0 \, k_{\lambda_i} \exp(-k_{\lambda_i} l_0) H\left(l_0 < L/\mu_0\right) = \int_0^{L/\mu_0} dl_0 \, k_{\lambda_i} \exp(-k_{\lambda_i} l_0) = 1 - \exp\left(-k_{\lambda_i} L/\mu_0\right) \text{ hence}$$

$$p = \int_{2\pi^+} d\omega_0 \, p_{\Omega_0}^{light}(\omega_0) \frac{C_A E_{A,\lambda_i}}{k_{\lambda_i}} \left[1 - \exp\left(-k_{\lambda_i} L/\mu_0\right)\right] \tag{B.2}$$

**B.1 Collimated emission**

When emission is collimated along a unique direction $\omega_i$, the distribution $p_{\Omega_0}^{light}(\omega_0)$ is a delta-Dirac distribution centered at $\omega_i$; therefore, the integral $\int_{2\pi^+} d\omega_0 \, p_{\Omega_0}^{light}(\omega_0)$ in Eq. B.2 vanishes and we obtain the following analytical solution for p:

$$p = \frac{C_A E_{A,\lambda_i}}{k_{\lambda_i}} \left[1 - \exp\left(-k_{\lambda_i} L/\mu_i\right)\right] \tag{B.3}$$

where $k_\lambda = C_A E_{A,\lambda} + C_B E_{B,\lambda}$ and $\mu_i$ is the cosine of the incident direction (for collimated normal incidence $\mu_i = 1$). This formulation of the probability p that a photon entering the medium is absorbed by the actinometer A reads as the products of the probability $\left[1-\exp\left(-k_{\lambda_i} L/\mu_i\right)\right]$ that a photon is absorbed within the medium, either by A or C, times the conditional probability $\dfrac{C_A E_{A,\lambda_i}}{k_{\lambda_i}}$ that a photon is absorbed by A knowing that the photon is absorbed. Note that $\left[1-\exp\left(-k_{\lambda_i} L/\mu_i\right)\right]$ can be read as 1 minus the probability that a photon entering the medium is transmitted, where the transmission probability is given by the standard exponential attenuation law.

## B.2 Lambertian emission

For a diffuse (i.e Lambertian) emitting light source, $p^{light}_{\Omega_0}(\omega_0) = \dfrac{\mu_0}{\pi}$; therefore Eq. B.2 gives

$$p = 2\pi \int_0^1 d\mu_0 \frac{\mu_0}{\pi} \frac{C_A E_{A,\lambda_i}}{k_{\lambda_i}}\left[1-\exp\left(-k_{\lambda_i} L/\mu_0\right)\right] = \frac{C_A E_{A,\lambda_i}}{k_{\lambda_i}}\int_0^1 d\mu_0\, 2\mu_0\left[1-\exp\left(-k_{\lambda_i} L/\mu_0\right)\right] \quad (B.4)$$

Using the exponential integral function $\mathrm{Ei}(x) = \int_{-x}^{+\infty} \dfrac{\exp(-t)}{t} dt$ leads to

$$p = \frac{C_A E_{A,\lambda_i}}{k_{\lambda_i}}\left[\exp\left(-k_{\lambda_i} L/\mu_0\right)\left((k_\lambda L)^2 \exp\left(k_{\lambda_i} L/\mu_0\right)\mathrm{Ei}\left(-k_{\lambda_i} L/\mu_0\right) + \mu_0\left(\mu_0\left(\exp\left(k_{\lambda_i} L/\mu_0\right)-1\right)+k_{\lambda_i} L\right)\right)\right]_0^1$$

where $\mathrm{Ei}(-x) \xrightarrow[x\to+\infty]{} 0$; therefore we obtain the following analytical solution for p:

$$p = \frac{C_A E_{A,\lambda_i}}{k_{\lambda_i}}\left[1+\exp(-k_{\lambda_i} L)(k_{\lambda_i} L - 1) + (k_{\lambda_i} L)^2 \mathrm{Ei}(-k_{\lambda_i} L)\right] \quad (B.5)$$

**Nomenclature**

| | |
|---|---|
| $\mathcal{A}_\lambda$: | Spectral local volumetric rate of radiant light energy, LVREA, ($\mu mol_{h\nu}.m^{-3}.s^{-1}.nm^{-1}$) |
| $\mathcal{A}$: | Local volumetric rate of radiant light energy, LVREA, ($\mu mol_{h\nu}.m^{-3}.s^{-1}$) |
| $a_{light}$: | Illuminated specific surface of the reactor ($m^2.m^{-3}$) |
| $C_i$: | Molar concentration of species i ($mol.m^{-3}$) |
| D: | Domain |
| e: | Optical thickness of the absorbing solution (-) |
| $E_{i,\lambda}$: | Molar absorption coefficient of species i at wavelength $\lambda$ ($m^2.mol^{-1}$) |
| $e_x$: | direction of the x axis in the cartesian coordinate system |
| H: | Heaviside function |
| $k_L a$: | Gas liquid volumetric mass transfer coefficient ($h^{-1}$) |
| l: | Absorption length (m) |
| L: | Geometric thickness (m) |
| $\ell$: | light path length (m) |
| N: | Total number of samples (-) |
| p: | Proportion of absorbed photons by the actinometer over emitted photons (-) |
| p: | Photon absorption probability in the reactor (-) |
| $p_\Lambda^{light}(\lambda)$: | Emission probability density function of the light source (-) |
| $\hat{p}_N$: | Estimator of the photon absorption probability |
| $q_\cap$: | Hemispherical photon flux density ($\mu mol.m^{-2}.s^{-1}$) |
| r: | random number for reflectivity comparison (-) |
| $r_i$: | Local volumetric reaction rate for species i ($mol.m^{-3}.s^{-1}$) |
| $S_{light}$: | Illuminated surface ($m^2$) |
| t: | Time (s) |
| V: | Volume of the reactive solution ($m^3$) |
| w: | Weight (-) |
| x: | Rectangular cartesian coordinate (m) |
| **x**: | Location defined by three coordinates ($x_1, x_2, x_3$) |
| **y**: | Location defined by three coordinates ($y_1, y_2, y_3$) |

**Other**

$\langle \ \rangle = \frac{1}{V} \iiint_V dV$ : Spatial averaging

**Greek letter**

| | |
|---|---|
| $\Phi$: | Quantum yield of the reaction (-) |
| $\theta$: | Angle between propagation direction $\omega$ and $e_x$ |
| $\omega$: | Direction of photon propagation (-) |
| $\tau$: | Reaction time corresponding to 99 % conversion in diffuse emission model (s) |
| $\rho$: | Reflectivity of the material (-) |
| $\lambda$: | Wavelength (nm) |
| $\mu_0$: | Cosine between the incident direction $\omega_0$ and the unit vector $e_x$ (-) |
| $\varepsilon_d$: | Dark fraction of the reactor (-) |
| $\sigma_N$: | Standard deviation (-) |

**Subscript**

| | |
|---|---|
| $\Omega$: | random direction for Monte Carlo algorithm |
| $\Lambda$: | random wavelength for Monte Carlo algorithm |
| $\lambda$: | Dependence on wavelength |
| 0: | initial value or first iteration |
| 1: | relative to first interaction |
| A: | relative to A chemical species or Reinecke salt |
| B: | relative to B chemical species or product $SCN^-$ |
| C: | relative to C chemical species or product $[Cr(NCS)_3H_2O(NH_3)_2]^-$ |
| i: | relative to wavelength |
| max: | maximum value |
| mean: | average over wavelength |
| min: | minimum value |
| out: | relative to the exit |

**Supercript**

| | |
|---|---|
| +: | relative to inner hemisphere (-) |